\documentclass[eqnl]{aa}         
\input{psfig.sty}

\def\simless{\mathbin{\lower 1pt\hbox
   {$\spose{\raise 5pt\hbox{$\char'074$}}\char'430$}}}
\def\simgreat{\mathbin{\lower 1pt\hbox
   {$\spose{\raise 5pt\hbox{$\char'076$}}\char'430$}}}
\def\simgreat{\gapp}
\def\simless{\lapp}
\def\lapp{\mathbin{\raise2pt \hbox{$<$} \hskip-9pt \lower4pt \hbox{$\sim$}}}
\def\gapp{\mathbin{\raise2pt \hbox{$>$} \hskip-9pt \lower4pt \hbox{$\sim$}}}

\begin{document}

%
\title{The hard X-ray emission from  the complex SNR MSH~15$-$52 
observed by BeppoSAX }

\author{T. Mineo\inst{1}, G. Cusumano\inst{1}, M.C. Maccarone\inst{1}, 
S. Massaglia\inst{2}, E. Massaro\inst{3,\dagger},   E. Trussoni\inst{4}}

\titlerunning{The hard X-ray emission from  the complex SNR MSH~15$-$52}
\authorrunning{T. Mineo et al.}

\institute{Istituto di Fisica Cosmica con Applicazioni
all'Informatica, CNR, Via Ugo La Malfa 153, I-90146, Palermo,
Italy \and
Dipartimento di Fisica Generale dell'Universit\`a,
Via Pietro Giuria 1, I-10125, Torino, Italy 
\and
Istituto di Astrofisica Spaziale, CNR, Via Fosso del Cavaliere,
I-00113, Roma, Italy \and
Osservatorio Astronomico di Torino, Strada Osservatorio 20,
I-10025, Pino Torinese, Italy}

\offprints{T. Mineo: mineo@ifcai.pa.cnr.it\\
$\dagger$ on leave from Universit\`a La Sapienza, Roma, Italy}

\date{Received: 8 August 2001; Accepted: 16 October 2001}

\abstract{ 
We present the results  of a BeppoSAX observation of
the Supernova Remnant MSH~15$-$52, associated with the pulsar PSR~B1509$-$58, 
 and discuss its  main morphological and spectroscopic properties 
in the  1.6--200 keV energy range (MECS and PDS instruments). The 
two main structures of the remnant,  the Southern Nebula, the 
plerion centered on the pulsar, and the Northern Nebula, are 
clearly visible in the MECS, with the former showing a much a harder 
spectrum. Furthermore, a diffuse extended emission  surrounds the 
whole remnant up to $\approx 17^{\prime}$ from the center. Non-thermal 
flux is  detected  in the PDS up to 200 keV  as well, and it appears that
also in this energy range the emission is not concentrated in the central
region around the pulsar. These data imply that the plerion 
extends up to a few tens of parsecs from the pulsar. } 

\maketitle


\section{Introduction}
The source MSH~15$-$52 (G320.4$-$1.2), associated with the pulsar
PSR~B1509$-$58, is one of the most  studied objects in the wide class of
 Supernova Remnants (SNR). The morphology of this remnant 
is quite complex: in the radio band it appears basically spherically
symmetric (size $\approx
30^{\prime}$), typical of shell-like SNR, with bright spots in the NW
and SE regions (see e.g. Milne et al. 1993, and references
therein). The NW region coincides with the IR and optical  nebula
RCW 89, where irregular filaments  are present, emitting in the
$H_{\alpha}$ line (Seward et al. 1983). Caswell et al.
(1975) estimated a distance of $\approx 4.2$ kpc and the 
 classical relationship between the surface brightness at 1 GHz 
and the linear diameter ($\Sigma-D$  relation) yields an  age
of $\sim 1 \div 2 \times 10^4$ years (Clark \& Caswell 1976).  At X-ray
energies, diffuse emission is detected from  the shell-like remnant with a
bright spot coincident with the NW zone. Centered at the pulsar position, a
plerion-like structure is also present with  size similar to that
of the northern region (see e.g. Trussoni et al. 1996, and
references therein).

The pulsar has a period of $\approx$ 150 ms  with one of the largest period
derivatives, $\approx 1.5 \times 10^{-12}$ s s$^{-1}$, and emits at radio
frequencies, in the soft and hard X-ray bands and at $\gamma$ energies 
(see e.g. Trussoni et al. 1990, Kawai et al. 1993, Matz et al. 1994, 
Greiveldinger et al. 1995, Rots et al. 1998).
The  spin-down age of $\approx$ 1700 years,  younger
than that estimated for the SNR, suggested possible associations
with the supernova that exploded in 185 AD (Thorsett 1992).
It was debated for a long time whether  PSR~B1509$-$58 and its surrounding 
plerion were components of the same system or just aligned along the line 
of sight. This apparent discrepancy can be reconciled assuming that the 
supernova exploded in an underdense cavity
(Bhattacharya 1990). This possibility has been confirmed by the
detailed radio observations of Gaensler et al. (1999).

Concerning the X-ray observations, several timing and
spectroscopic data for the pulsar are available from the soft to the 
hard band. 
Spectroscopic and morphological data are necessary for the analysis of
the properties of the remnant. So far, information
 on the structure of the remnant are
available the in the low energy  X-ray band (Einstein and ROSAT), 
up to 10 keV by the observations of  EXOSAT and ASCA and at very high 
energies (up to 250 keV) by the Ginga and RXTE missions.

In  order to  better understand the physics of these peculiar objects we
observed MSH~15$-$52 and PSR~B1509$-$58 with BeppoSAX in February 1998.
The instrumentation on board this satellite provides
spectroscopic analysis in a very wide X-ray band
(0.1--200 keV) with a spatial resolution  of $\approx 1.5^{\prime}$,
up to energies of 10 keV. Here we present
the results of the observation of the extended nebula, while the data
on the pulsar emission are discussed in a separate paper (Cusumano et al.
2001). The main result of the present analysis is the discovery that 
the hard emission up to about 200 keV very likely originates from the 
extended nebula and not only from the inner plerion around the pulsar.
In  Sect. 2 we summarize the main X-ray properties of
MSH~15$-$52 as obtained in previous missions, in Sect. 3 we outline the 
details of the observations and of the procedure for data reduction, in 
Sect. 4 we present the results of our analysis, while their astrophysical 
implications are discussed in Sect. 5.

\section{The X-ray structure of MSH~15$-$52 from previous observations}

 From the images of the remnant  obtained by ROSAT in the soft X-ray band 
(Fig. 1) we see the complex structure of MSH~15$-$52 (for details see
Greiveldinger et al. 1995, Trussoni et al. 1996, Brazier \& Becker
1997). Around the pulsar position there is the plerion with quite
uniform emission ({\it Southern Nebula}) and a cross-like
shape, with  the longest arms in the SE-NW direction  (dimensions $\approx
10^{\prime} \times 6^{\prime}$). The northern
arm merges into the {\it Northern Nebula} (coinciding with RCW
89), with dimensions $\approx
9^{\prime} \times 4^{\prime}$ at a distance of $\approx 7^{\prime}$ 
from the pulsar. In the HRI image the  Northern Nebula shows an inhomogeneous morphology 
with several knots of emission (Brazier \& Becker 1997).
The opposite arm of the  Southern Nebula ends in a
curved tailed  component  with emitting blobs in the SE
direction. All these structures are embedded in a diffuse and
symmetric region,  approximately centered at the pulsar 
position ({\it Central
 Diffuse
Nebula}), of radius $\approx 15^{\prime}$. Outside  this region
 there  is a  weak
emission extended in the SE direction ({\it Southern Extended
Nebula})  and coincident with the radio source G320.6-1.6 (Milne
et al. 1993).

\begin{figure}
\centerline{
\vbox{
\psfig{figure=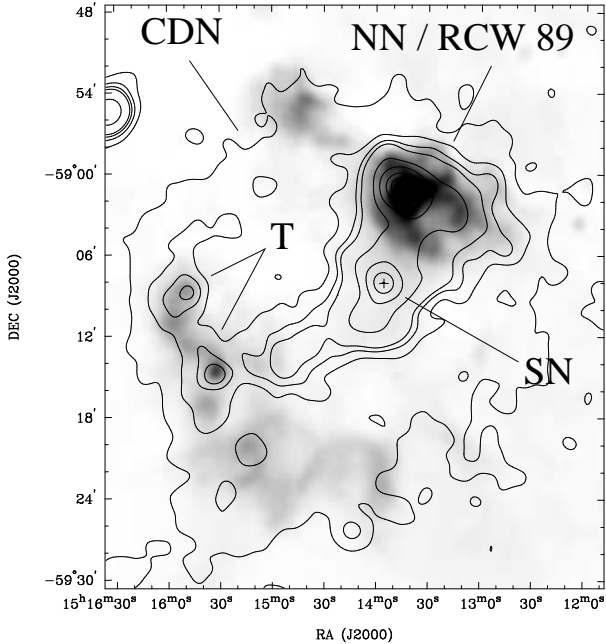,width=8cm,angle=-90,lip=}
}} 
\caption{Radio image of MSH~15$-$52 (36 cm) with the contour plot  
from the  ROSAT PSPC observation (0.1--2.4 keV). The cross marks
the position of PSR~B1509$-$58 (Gaensler et al. 1999).  NN and
 SN indicate the Northern and
Southern Nebulae respectively, T is the tail component and CDN the 
Central Diffuse Nebula.}
\end{figure}

The X-ray flux from the  Southern Nebula in the ROSAT band is consistent 
with   a power law spectrum, 
with  photon index $\Gamma \approx 2$ and  column density $N_{\rm H}$ 
in the range  $7 \div 9 
\times 10^{21}$ cm$^{-2}$. 
More complex is the case of the  Northern Nebula: its  PSPC
spectrum fits with a thermal 
model with $kT \approx 0.4$ keV but with a column density about two 
times larger than in the  Southern Nebula. 
However, ASCA observations of the  Northern Nebula (Tamura et al. 1996) showed that 
the emission is consistent with a composite model: a thermal spectrum 
(in agreement with the PSPC results) plus  a power law component, 
with $N_{\rm H} \approx 6 \times 10^{21}$ cm$^{-2}$ and $\Gamma 
\approx 2$, similarly with the values found by the PSPC for the  Southern Nebula.
These results may be reasonably interpreted if the two arms in the NW-SE
directions are opposite outflows emerging from the pulsar, while the shorter
arms in the NE-SE direction could be related to an equatorial torus
surrounding the pulsar, as proposed for the Crab Nebula (Hester et al. 1995, 
Brazier \& Becker 1997). Collimated winds accelerated from neutron stars 
have been proposed also to interpret the X-ray morphology of the Vela 
SNR (Markwardt \& \"Ogelman 1995), and were recently detected in a Chandra
observation of the pulsar in the Crab Nebula (Weisskopf et al. 2000)
and in the Vela SNR (Pavlov et al. 2001).

This picture of the remnant is further supported by the 
detailed radio observations of Gaensler et al. (1999) who  suggested a
strict interaction of the pulsar with the surrounding remnant.
In particular, the X-ray knots detected in
the  Northern Nebula and in the  tail coincide with analogous compact radio
structures. These are interpreted as regions where the collimated
outflows, originating from the pulsar, strongly interact with the
environment. The positions of these radio/X-ray blobs in the  Northern Nebula
suggest that the direction of the jet axis changes with the
time, following a sort of precession path.  These radio data
strongly support  that, in spite of the apparent different ages,
MSH~15$-$52 and PSR~B1509$-$58 originated in the same event (conversely
it is very likely that G320.6$-$016, coincident with the  Southern Extended Nebula, is an
older SNR along the same line of sight).

In harder X-ray bands, apart from ASCA, we have data only from non imaging 
instruments. The results from the observations of 
EXOSAT/ME (up to 11 keV, Trussoni et al. 1990), Ginga (up to
60 keV, Kawai et al. 1993) and RXTE (up to 250 keV, Marsden et al. 1997) 
confirm that also at these energies the spectrum  follows a power law  
with a photon index
$\approx 2 - 2.2$, but the flux is about three times higher than
deduced by the PSPC for the Southern Nebula.  Furthermore, Marsden et al. (1997) 
found that a narrow iron line  at $6.7$ keV significantly improved 
 their fit.

\section{BeppoSAX observation and data analysis}

\begin{figure}
\centerline{
\vbox{
\psfig{figure=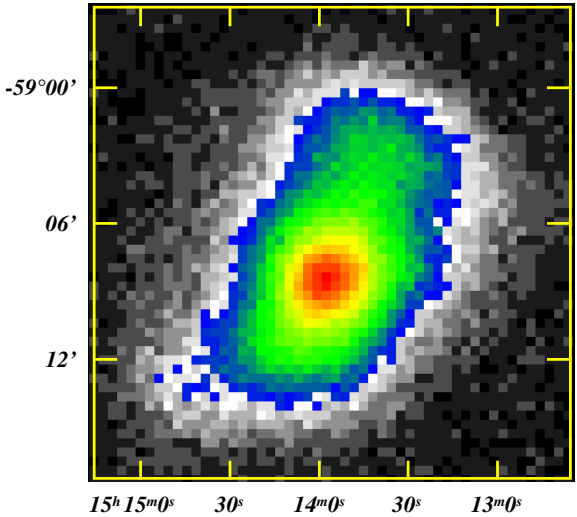,width=8cm,clip=}
}}
\caption{Image of the complex SNR MSH~15$-$52 obtained by the MECS 
observation  (1.6--10 keV).  
Spatial  rebinning to a pixel size of 24 $\times$ 24 arcsec was applied.
The position of the pulsar PSR~B1509$-$58 coincides with the bright 
red spot at the center; this maximum corresponds to 1006 counts per pixel. 
}
\end{figure}

The BeppoSAX observation of MSH~15$-$52 was performed on 9--11 February
1998 for a total exposure of 31\,008 s in the LECS, 82\,483 s in
the MECS and  34\,493 s in the PDS (for details on the BeppoSAX
instrumentation see Boella et al. 1997). 
Standard procedures and selection criteria were applied to
the data using the SAXDAS v.2.0.0 package.

\begin{figure}
\label{fig2}
\centerline{\psfig{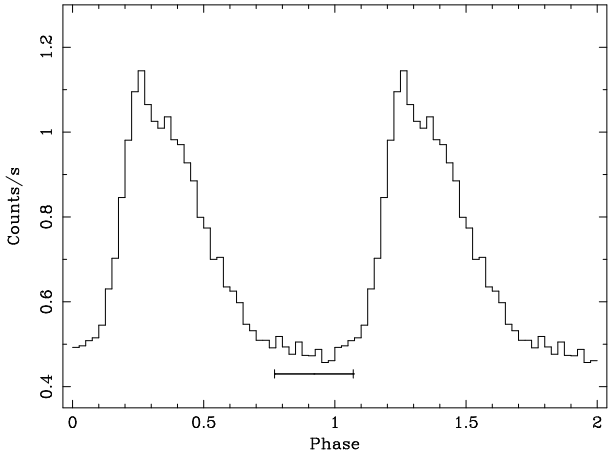}}
\caption{ Pulse profile of PSR~B1509$-$58 in the 1.6--10 keV energy band
(MECS). The horizontal line corresponds to the phase interval used in the
spectral analysis of the nebula.}
\end{figure}

A raw image  of the remnant as it appears in the 
MECS (units 2 and 3) integrated over the entire energy band (1.6--10 keV) 
is shown in Fig. 2. The elongated structure in the SE-NW direction is 
clearly visible but the limited resolution does not allow one to resolve 
the minor substructures. It is possible
to  improve the image quality, up to the MECS spatial resolution, 
by applying a deconvolution algorithm based on the Lucy (1974) method
to each  monochromatic image and using the proper local Point Spread 
Function (PSF).
The adopted procedure is  described in  detail in Maccarone et al. (2001). 
For each energy channel MECS images were first rebinned to a pixel size
of 24 $\times$ 24 arcsec to improve the statistics, and then deconvolved 
with the proper monochromatic PSF.
According to Maccarone et al. (2001) we used   300 iterations 
which guarantee a good convergence at all the energies. 
Final processed images were then added 
together in selected energy  bands.  In this analysis
the pulsar contribution has not been eliminated 
to avoid  strong reduction of the statistics. 

To perform the spectral analysis of the nebular emission of the  Southern Nebula
 we  excluded the contribution from the pulsar.  
Phase histograms, for each energy channel of the 
NFIs instruments, have been obtained folding  the selected UTC 
arrival times, converted  to the Solar  System Barycenter,  with 
the pulsar ephemeris (see Cusumano et al. 2001 for detail). Fig. 3 shows 
the resulting pulsed profile
for the  energy band 1.6--10 keV; the nebular emission has been extracted
from the off-pulse phase interval
(0.77--1.07) where we assume that the pulsar emission is not dominant.
Spectral analysis was performed with the XSPEC package
(ver. 11.0) after rebinning the counts in order to have at least 20
events in each energy channel. 
As the source is extended, a suitable auxiliary matrix must be
created to correct for the vignetting, that is relevant at distances
$\simgreat \ 4^{\prime}$ from the center of the FOV (Fiore et al. 1999).
This can be performed for the MECS data using the SAXDAS {\it effarea}
command, but the procedure is 
suitable only for sources with symmetric azimuthal intensity
distribution  (for details see Molendi 1998, D'Acri et al. 1998).   
We then produced an ad-hoc matrix  computing the vignetting correction
according to the asymmetric spatial distribution  of  MSH 15$-$52.   
We then verified that using such auxiliary matrix the resulting best fit 
values of the spectral indices differ from those obtained with the 
standard matrices  by less than 
$\approx  2 \sigma$ while the normalizations show a discrepancy of
$\approx$ 30\%.  
 No analogous procedure is  available at the moment for the LECS 
observations of extended targets,  therefore we have not included these 
data in our analysis.   

Spectral fits were first performed separately for the MECS and 
PDS data and after a combined fit of both data sets was also considered. 
The spectrum  of the background for the MECS has been
extracted from blank fields, while the PDS one is simultaneously monitored
by rocking the collimator. 
The uncertainties  reported in the following are at  1 standard deviation
 for one interesting parameter.

\section{Results}

\begin{figure}
\label{fig4} 
\centerline{ \hbox {
\psfig{figure=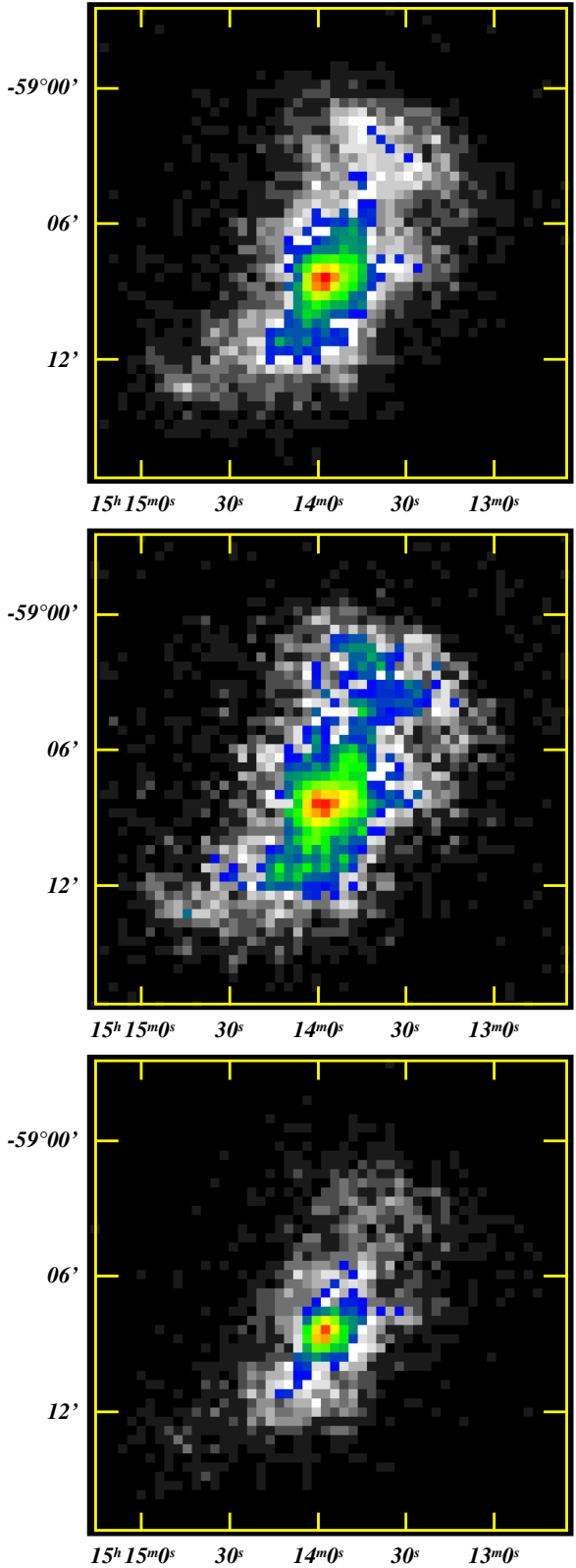,width=8cm,clip=} }} 
\caption{Deconvolved images of the SNR MSH~15$-$52 in the three energy 
bands: 1.6--10 keV (upper panel), 1.6--4 keV (central panel), and 4--10 
keV (bottom panel). 
The maximum count levels are 4152, 1584, 2571 counts per pixel, 
respectively.  
}
\end{figure}

\begin{figure}
\label{fig5} 
\centerline{ \hbox {
\psfig{figure=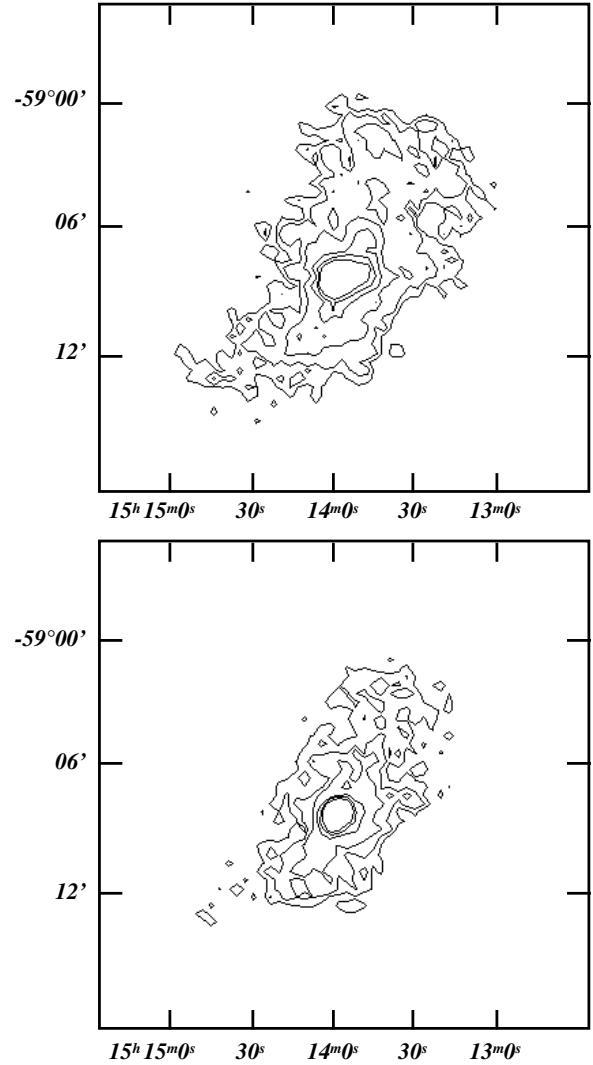,width=8cm,clip=} }} 
\caption{Contour plots of the soft ( 1.6--4 keV, upper panel) and hard 
( 4--10 keV, lower panel) deconvolved images of the 
Fig. 4. The contour levels correspond to the fractions of  1.00, 0.75, 
0.50, 0.25, 0.10, and 0.05 of the mean central intensity after the
subtraction of the contribution from  PSR~B1509$-$58.
}
\end{figure}

In this section we discuss first the remnant morphology in the intermediate
band (1.6--10 keV) as obtained from the deconvolved MECS data, then we
address  the spectroscopic properties obtained from the MECS and PDS 
observations.

\subsection{Morphology}

As expected from previous observations (see Sect. 2),
the MECS image of this plerionic system shows an extended emission elongated
 at about the Position Angle $-$20$^{\circ}$,  with a bright point source
in the center corresponding to the pulsar position. 
The deconvolved images in three energy bands are shown in Fig. 4: 1.6--10 keV 
(upper panel), soft X-ray band 1.6--4 keV (central panel), hard X-ray band
4--10 keV (upper panel).  

The contour plots of the last two images are also 
shown in Fig. 5.    The contour levels are relative 
to the mean central intensity measured after  subtraction
of the contribution from the pulsar.
In all these deconvolved images it is possible to recognize the major 
structures already known from ROSAT (Fig. 1). In particular, 
the much softer emission from the  Northern Nebula is evident, that almost disappears for photon 
energies higher than 4 keV.
In the harder band we notice the dominant central symmetric region,
related to the pulsar emission, and the elongated structures towards the
 Northern Nebula and the tail region, respectively. This confirms the ASCA results 
(Tamura et al. 1996), that showed a highly  asymmetric structure of 
the MSH 15--52 up to X-ray energies $\simgreat $ 10 keV. 
The angular dimensions of the  Southern Nebula in the higher energy range are at least 
 $4^{\prime} \times 8.5^{\prime}$ (at the 10\% level of the nebular
central intensity), 
corresponding to about 5 $\times$ 10 pc. 
 
 The soft spectrum of the  Northern Nebula is also confirmed by the hardness ratio
of these two images using pixel size of $2^{\prime} \times 2^{\prime}$ 
to improve the statistics (Fig. 6).
The hardest value in the center of the nebula
corresponds to the pulsar position, as expected.

We remark finally that due to the smaller FOV of the MECS, from the image 
we cannot disentangle the presence of the large  Central Diffuse Nebula detected in the ROSAT
observation.

\subsection{The spectral properties of the plerionic emission}

In the following we first analyze separately the data from the MECS 
and the PDS and after we present the combined analysis.

\subsubsection{MECS data}
Due to the complex structure of the remnant, we performed
spectral fits to flux measured in the MECS 
from three  different regions.  The first two  are centered at the 
pulsar position with radii $4^{\prime}$, corresponding to the central 
region of the Southern Nebula, and $17^{\prime}$,
to include the extended emission excluding however a sector 
containing the Northern Nebula;
the third region is centered on the Northern Nebula with
radius $3^{\prime}$. For the first two, events from  the off-pulse phase 
interval are considered, while, for the third one, no phase selection was 
performed because the pulsar contamination at such a distance was estimated 
around 3\%.
Spectral fits were computed for a power-law distribution and
the resulting parameter values are reported in Table 1.
From the fit to the  Southern Nebula emission ($4^{\prime}$) we have 
deduced for the column density $N_{\rm H} = (8.0 \pm 1.5) \times 10^{21}$ 
cm$^{-2}$, consistent within 1 $\sigma$ with the value deduced from 
Cusumano et al. (2001), $9.1 \times 10^{21}$ cm$^{-2}$, that we have 
adopted in our analysis.
\vskip 0.1 true cm
\noindent
{\it Southern Nebula and Central Diffuse Nebula.} 
The spectral slopes (Table 1) are much steeper than the pulsar one
that was found to be equal to $\approx 1.2$.
The flux from the larger area, including also the  Central Diffuse Nebula contribution,
has a slightly steeper spectral distribution slope, while the 2--10 keV energy 
flux is approximately two times larger. If we include in this fit also
the sector with  Northern Nebula, the spectrum is slightly steeper
($\Gamma = 2.08 \pm 0.01$) with a  the total 2--10 keV flux 
of $ 1.2 \times 10^{-10}$ erg cm$^{-2}$ s$^{-1}$, i.e. 
$\approx 1.3$ times higher than in the previous case with the  Northern Nebula excluded.  
           
 We have searched for  possible line emissions at 6.4 and 6.7 keV in 
the MECS spectrum extracted over the $17^{\prime}$ radius region 
and found no evidence of such features.  The 6.7 keV upper limit at 90\% 
confidence level, of 1.2 $\times$ 10$^{-4}$ photons cm$^{-2}$ s$^{-1}$, 
 about 30\% less than the line flux derived from 
RXTE data  (Marsden et al. 1997), is 
 compatible with the Galactic ridge emission (Koyama et al. 1989).   
\vskip 0.1 true cm
\noindent
{\it Northern Nebula.} 
It is known from the ASCA data (Tamura et al. 1996) that 
a thermal and  a power-law components contribute to the total flux, 
with the former that largely prevails below 3 keV. 
In order to avoid the contamination from thermal photons, we fitted 
the MECS data for energies $\geq 4$ keV  and found a
photon index of 2.10,  steeper than that found for  Southern Nebula
(Table 1).
We estimate that this power-law component contributes for $\approx 25 \%$
to the total flux of the remnant in the 2--10 keV band, as expected
from the previous results on the Southern Nebula and Central Diffuse Nebula.

\subsubsection{PDS data}
The power-law fit of the unpulsed high energy flux detected in the PDS 
gave a poorly constrained photon index and a total flux of $ 1.2 \times 
10^{-10}$ erg cm$^{-2}$ s$^{-1}$ in the  20--200 keV energy range
(see Table 2). 
The slope is consistent with the extrapolation of the spectrum at lower 
energies obtained with the MECS, implying that energy losses do not 
crucially affect the emitting electrons up to energies of 200 keV. 
The lack of imaging information and the not enough constrained  evaluation  
of the energy flux do not allow a definite answer to the question 
concerning the spatial extent of the region emitting at these energies. 

\subsubsection{Combined MECS and PDS data}

A spectral analysis of the combined data sets from the MECS (with 
extraction radii of $4^\prime$ and $17^{\prime}$) and the PDS can provide 
information on the size of the region for the hard X-ray emission.
Note that now the MECS spectrum obtained with an extraction radius 
of $17^{\prime}$ includes the  Northern Nebula contribution.  
 When evaluating these fits, we considered as a free parameter the 
normalization factor $f_{\rm ic}$ by which the MECS spectrum 
must be multiplied to match that from PDS. According to the in-flight 
calibrations with well-known sources, it is expected that 
$0.77 \la f_{\rm ic} \la 0.95$ (Fiore et al. 1999), but this 
 range applies only for point-like sources, while its value has not been 
calibrated for  significantly extended sources, as in the present case.
From the results reported in Table 2, we see that the values of this 
parameter are very different if the MECS data from the small or large 
region are considered.  
However, near the center of the FOV the vignetting 
effects are not relevant, as confirmed by our tests in Sec. 3,  so 
that the  Southern Nebula can be quite reasonably 
considered point-like in the MECS. 
Considering now the linear off-axis response of the PDS, if the region
of emission of the hard photons is comparable (or smaller) than the  Southern Nebula, 
we can assume again that it is `seen' basically point-like in the detector. 
Then, considering the hard component as the high energy tail 
of the emission detected in the MECS, and originating from the 
inner region of the plerion, we would expect  $f_{\rm ic}$ to lie
in the range 
given above, from the fit of the MECS ($4^{\prime}$) + PDS 
spectra. In contrast, we find $f_{\rm ic} \approx 1.35$  
(see Table 2),
  higher 
than the expected  point source upper  limit: 
 this result implies that
the MECS flux must be increased by 35\% instead of being reduced  about
15\%, as in the case of a point source. 
This discrepancy may be reasonably related to the fact that the region 
emitting in the 20--200 keV is more extended than the  Southern Nebula 
itself.

\begin{table*}
\caption{Power-law best fit parameters of the plerionic spectrum: MECS 
(1.6--10 keV)}
\begin{tabular}{llll}
\hline
Parameter$^{\rm a}$     & MECS ( SN$^{\rm e}$, $4^{\prime}$) & MECS$^{\rm
b}$ ($17^{\prime}$) & MECS ( NN, $3^{\prime}$)  \\ \hline
Photon index ($\Gamma$) & $1.90 \pm 0.02$   & $2.04 \pm 0.01$  & 2.10$\pm$0.06    \\
Flux at 1 keV$^{\rm c}$ & $1.56  \pm 0.05$  & $3.68 \pm 0.07$  & 1.38$\pm$0.14    \\ 
Energy flux $^{\rm d}$  & 0.47              & 0.89             & 0.31              \\ 
$\chi^2$ (dof)          & 0.99 (91)         & 0.96 (91)       & 0.99 (55)        \\
\hline
\end{tabular}
\vskip 0.15 true cm
$^{\rm a}$ $N_{\rm H} = 9.1 \times 10^{21}$ cm$^{-2}$
\vskip 0.1 true cm
$^{\rm b}$ excluding a sector with the  Northern Nebula (NN)
\vskip 0.1 true cm
$^{\rm c}$ $\times 10^{-2}$ photons cm$^{-2}$ s$^{-1}$ keV$^{-1}$
\vskip 0.1 true cm
$^{\rm d}$ unabsorbed $\times 10^{-10}$ ergs cm$^{-2}$ s$^{-1}$, 2--10 keV (MECS)   
\vskip 0.1 true cm
 $^{\rm e}$ SN: Southern Nebula 
\end{table*}

\begin{table*}
\caption{Power-law best fit parameters of the plerionic spectrum: 
PDS (20--200 keV) and MECS + PDS}
\begin{tabular}{llll}
\hline
Parameter$^{\rm a}$    & PDS & MECS (SN, $4^{\prime}$) + PDS  &  MECS$^{\rm b}$  ($17^{\prime}$) + PDS   \\
\hline
Photon index ($\Gamma$)  & $2.1 \pm 0.2$  & $1.90 \pm 0.02$   &   2.08 $\pm$ 0.01       \\
Flux at 1 keV$^{\rm c}$  & $4.0 \pm 2.5 $ & $1.57 \pm 0.0.05$ &   5.16  $\pm$ 0.09      \\ 
Energy flux $^{\rm d}$   & 1.20            &                   &   1.37                  \\ 
f$_{\rm ic}$             &                & $1.35 \pm 0.12$   &   0.73 $\pm$ 0.06       \\ 
$\chi^2$ (dof)           &  0.68 (6)      &  0.87 (98)        &   1.04 (98)             \\
\hline
\end{tabular}
\vskip 0.15 true cm
$^{\rm a}$ $N_{\rm H} = 9.1 \times 10^{21}$ cm$^{-2}$
\vskip 0.1 true cm
$^{\rm b}$ including the  Northern Nebula
\vskip 0.1 true cm
$^{\rm c}$ $\times 10^{-2}$ photons cm$^{-2}$ s$^{-1}$ keV$^{-1}$
\vskip 0.1 true cm
$^{\rm d}$ $\times 10^{-10}$ ergs cm$^{-2}$ s$^{-1}$, 20--200 keV (PDS)              
\end{table*}

\begin{figure}
\label{fig6} 
\centerline{ \vbox {
\psfig{figure=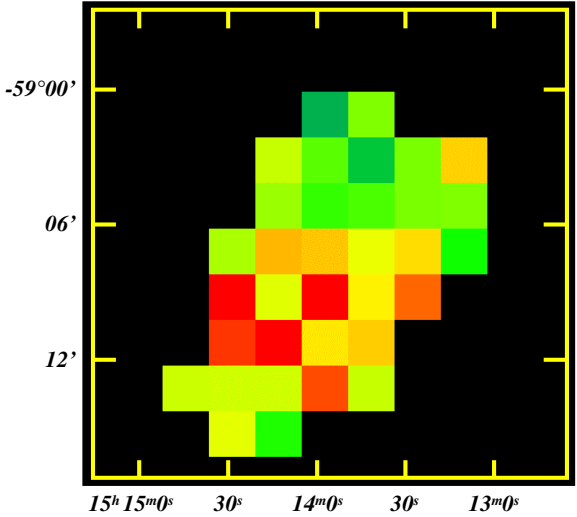,width=8.0cm,clip=} }} 
\caption{Hardness
ratio computed between the energy bands: 4--10 keV, 1.6--4 keV.
From Fig. 4 it is possible to see the location of the regions where the
hardness ratio has been computed.}
\end{figure}

\begin{figure}
\label{fig7}
\centerline{
\hbox {
\psfig{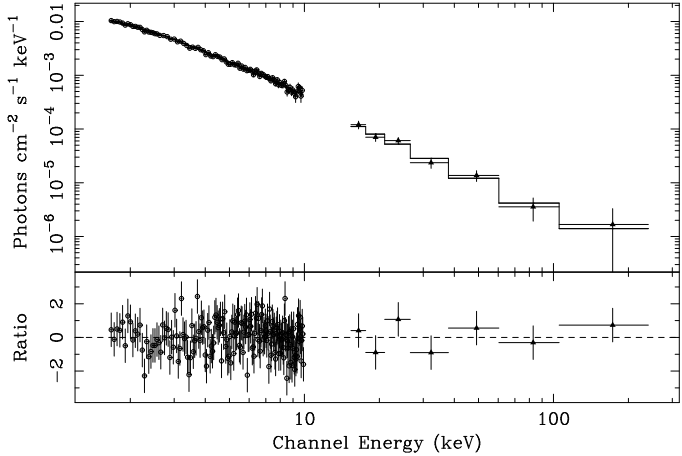}
}} 
\caption{MECS ($17^{\prime}$ and including the  Northern Nebula) and PDS spectrum of 
the whole nebula
}
\end{figure}

\section{Discussion}

The main results of this analysis concern the characteristics of the hard
X-ray flux from the remnant: it turns out from the MECS + PDS 
data that the non-thermal unpulsed emission extends over the range 
1.6--200 keV without any evidence of a break in the spectral slope. 
As far as the spatial emission distribution is concerned, in the 
range of the MECS it extends, without a strong 
break in the spectrum, up to  $\simgreat \, 17^{\prime}$ (i.e. $\simgreat 
\, \ 20$ pc) from the compact plerion that surrounds the pulsar. Furthermore
there are some clues that also the emission up to $\approx 200$ keV may 
 originate  far from the central region. 
 
Before  discussing the implications of these data on the structure of MSH
15-52  one has to consider that X-ray and $\gamma$-ray emission has been detected 
from the galactic ridge, where the remnant is located. In the energy range 
2--11 keV the average flux (consistent with a thermal continuum with 
temperature $\approx 9$ keV) is $8.0 \times 10^{-8}$ erg cm$^{-2}$ 
s$^{-1}$ sr$^{-1}$ (Koyama et al. 1986), which is $\approx 20$ times 
less than the flux (for arcmin$^2$) detected in the MECS. 
Regarding the hard range (20--200 keV), a diffuse non-thermal emission 
from the galactic ridge has been detected, but with a much steeper slope 
than we obtain in the PDS ($\Gamma \approx 2.7$, Skibo et al. 1997). 
Since no spurious hard source has been found in the field of view of 
the MECS, it is very reasonable that the hard spectrum is associated with 
the remnant and originates from a region more extended than the  Southern Nebula.  

The values of the spectral parameters are basically consistent with those 
derived, in the same range of energy, from the RXTE observation by Marsden 
et al. (1997) whose data were fitted by a spectrum with a slightly softer 
index ($\Gamma = 2.21$). However, in contrast with their conclusion,  
from the above arguments it is reasonable to conclude that the emission 
does not originate only from the inner plerion. Concerning  the Northern Nebula,  the slope of the spectrum is consistent
with that deduced from ASCA  within 1$\sigma$,
while the flux at 1 keV is $\approx 2$ times higher. Considering 
however the statistical errors the lower (in the MECS) and upper limits 
(in ASCA) of the fluxes differ by a factor $\approx 1.5$ and $\approx 1.2$
at 1$\sigma$ and 2$\sigma$ levels, respectively.

The total luminosity of the remnant in the  2--10 keV energy range is
$2.6 \times 10^{35}$ erg s$^{-1}$ (distance = 4.2 kpc) to which the 
 Southern  and  Northern Nebulae 
contribute by $\approx 40 \%$ and $\approx 25 \%$, respectively. The 
remaining fraction of $\approx 35 \%$ comes from the extended region which  
is 
reasonable to associate  with the  Central Diffuse Nebula detected from 
ROSAT in the soft X-ray band.  If the whole spectrum does not change up 
to $\sim$ 200 keV, the  Northern Nebula and 
the Central Diffuse Nebula should contribute, by comparable  
amounts, to more than half of the flux detected in the PDS. 

Hard X-ray emission is quite common in  plerions and shell-like
supernova remnants, which means that electrons of very
high energy must be present. For the latter objects 
(like e.g. Cas A and SNR 1006) it is proposed that these particles are
created  through both  the initial shock wave and local stochastic acceleration
processes, and  also a non thermal bremsstrahlung could be present
(Reynolds 1996, Favata et al. 1997). In plerions conversely high energy
electrons may originate from the pulsar and further be stochastically 
accelerated  in the shock where the relativistic pulsar wind encounters
the inner region of the remnant (Rees \& Gunn 1974, 
Kennel \& Coroniti 1984). A classical example of this picture is the Crab 
 Nebula, where the shock should lie at  $\sim 0.1$ pc from the pulsar,
however  the presence of a jet implies  that
the interaction of the pulsar outflow with the environment is anisotropic 
and very complex. 
We can reasonably imagine that a similar
structure exists in the inner region of the Southern Nebula, with emission at high X-ray 
energies (notice its slightly flatter spectrum). The major problem in our case 
is the extension of the remnant, with  $\approx 60 \%$ of the flux 
(up to 200 keV) originating several pc away from the inner plerion. 

It is extremely difficult to describe the complicated high energy structure 
of MSH~15$-$52 through a detailed dynamic
model of a pulsar relativistic wind interacting with a remnant. 
However we can try to see  whether the radiating particles in the 
Central Diffuse Nebula may
directly originate from the pulsar. 
In the simplest hypothesis, we 
can compare the Lorentz factor $\gamma$ of an electron which 
is synchrotron emitting 
a photon of energy $E$ in a magnetic field $B$:
$$
\gamma \approx  4.5 \times 10^8 (E_{\rm keV}/B_{\rm \mu G})^{1/2}\,\,,
\eqno(1)
$$
\noindent
($E_{\rm keV} = E/1$ keV and $B_{\rm \mu G} = B/10^{-6}$ G),
with that of an electron originating 
from the outer gap region (Zhang \& Cheng 1997, eq. 26):
$$
\gamma_{\rm og} \approx 3 \times 10^7 f^{1/2} (B_{12} 
R^3_{\star\,6}/ P)^{1/4}\,\,,
\eqno(2)
$$
\noindent
where $R_{\star\,6}$ is the radius of the pulsar in units of $10^{6}$  cm,
 $P$ is the period  in s,
$B_{12}$ the surface magnetic field in units of $10^{12}$ G and 
$f$ a factor $\leq 1$
that takes into account the volume fraction of the outer gap in the 
magnetosphere. Assuming as reasonable value for the average nebular magnetic 
field  $B=10~\mu G$  (Seward et al. 1984) in Eq. 1 and the parameters of 
PSR~B1509$-$58 in Eq. 2 ($P=0.15$ s, 
$B_{12} = 15.5$ and $R_{\star\,6}=1$), even in the extreme case of $f=1$ 
we find that $\gamma_{\rm og}$ is less by a factor $\approx 5$ and  
$\approx$ 15 than required for the 
X-ray emission in the extended region at 10 and 100 keV, respectively. 

The above reasoning is not entirely model-free, but relies upon theories of
particle acceleration at the outer gap. However we can  also draw similar
conclusions if we postulate that an alternative process produces
very high energy particles around the pulsar magnetosphere or in the
inner region of the plerion. In fact 
a relativistic electron in a mean magnetic field,  with 
a  cooling  time scale
(neglecting other radiative losses) : 
$$ \tau = \gamma/ |\dot \gamma|= 1.7 \times 10^{12} (E_{\rm keV}  
B^3_{\rm \mu G})^{- 1/2}\;\; {\rm s}\,\,,
\eqno(3) 
$$
\noindent
covers a distance:
$$
R \approx \kappa c \tau = 1.6 \times 10^4 (E_{\rm keV} 
  B^3_{\rm \mu G})^{-1/2} \;\; {\rm pc}\,\,,
\eqno(4)
$$
\noindent
where through the factor $\kappa \leq 1$ we take into account that the
particle  cannot follow a pure radial propagation outwards. Practically
we can consider $\kappa \sim
R_{\rm c}/D$, i.e. the ratio of the electron gyroradius to the linear size of
the remnant.  With  $B=10 \mu G$ the maximum distance attained by the
particle is  $R \approx 50 \kappa$ pc for $E = 100$ keV photons.  Consistently
with the extent of the Central Diffuse Nebula, one would require $\kappa
\sim 0.5$, while the actual value of the electron gyroradius for a 100 keV
emitting electron is $\sim 2 \times 10^{11}$ cm, implying a value of $\kappa
\lapp 10^{-8}$.

Then, excluding that the Central Diffuse Nebula is 
an old SNR on the line of sight and just overlapped and centered on the
pulsar,  it appears that in this remnant effective reacceleration 
processes must be effective far from the central plerion. 

Coming to the Northern Nebula,  the limited 
imaging capabilities of the MECS do not allow us to check whether this power law
emission is related to the jet emerging from the pulsar (as proposed by
Tamura et al. 1996) or to some region of local particle acceleration. 
In the former case in the collimated wind highly relativistic electrons must 
be present, and we find again the problem of the origin of such particles 
discussed above. If most of the hard emission is related to local
acceleration (e.g. shocks), we could argue that a fraction  of high energy
particles might diffuse away from the  Northern Nebula 
and contribute to the emission of the  Central Diffuse Nebula.
The detailed radio images of the Northern Nebula obtained by 
Gaensler et al. (1998) show 
discrete highly polarized spots coincident with the X-rays knots detected
in the ROSAT/HRI image (Brazier \& Becker 1997). These structures, whose 
emission is consistent with a single power law from radio to X-ray energies,
could be interpreted as shocks due to the interaction of a precessing jet
with the environment.  It is possible to see that the total X-ray 
emission from these knots is about twice that deduced from the
ASCA observations, but more consistent with the MECS results. 
In conclusion, the physical conditions of the emitting
plasma in the Northern Nebula are quite complex, and only detailed 
observations from
the new generation satellites will contribute new insights 
on this point.  

\begin{acknowledgements}
The authors are grateful to F. Fiore for his information on the analysis 
of extended images.
ET and SM acknowledge financial support from the Agenzia Spaziale Italiana 
through the grant ASI ARS 99 15. 
EM acknowledges financial support from the Italian Ministry of University
and Scientific Research through the grant Cofin 99-02-02. 
\end{acknowledgements}

\end{document}